\documentclass[sigconf, natbib=false]{acmart}
\usepackage[style=ACM-Reference-Format, maxbibnames=3, backend=bibtex, sorting=none]{biblatex}

\usepackage{pifont}
\usepackage{amsmath}
\usepackage{balance}
\usepackage{threeparttable}
\usepackage{xcolor}
\usepackage{soul}
\usepackage{subcaption}
\usepackage[linesnumbered, ruled, vlined]{algorithm2e}
\let\oldnl\nl
\newcommand{\nonl}{\renewcommand{\nl}{\let\nl\oldnl}}
\usepackage{comment}

\usepackage{array}
\usepackage{algcompatible}

\usepackage{multirow}
\usepackage{graphicx}
\usepackage{color, soul}
\usepackage{color, colortbl}

\usepackage{makecell}

\usepackage{breqn}

\usepackage{multirow}
\usepackage{breqn}
\AtBeginDocument{%
  \providecommand\BibTeX{{%
    \normalfont B\kern-0.5em{\scshape i\kern-0.25em b}\kern-0.8em\TeX}}}

\setcopyright{none}
\renewcommand\footnotetextcopyrightpermission[1]{}

\copyrightyear{2023}
\acmYear{2023}
\setcopyright{acmlicensed}\acmConference[GLSVLSI '23]{Proceedings of the Great Lakes Symposium on VLSI 2023}{June 5--7, 2023}{Knoxville, TN, USA}
\acmBooktitle{Proceedings of the Great Lakes Symposium on VLSI 2023 (GLSVLSI '23), June 5--7, 2023, Knoxville, TN, USA}
\acmPrice{15.00}
\acmDOI{10.1145/3583781.3590264}
\acmISBN{979-8-4007-0125-2/23/06}

\usepackage{tikz,tkz-euclide,pgfplots}
\tikzstyle{roundnode} = [circle, fill=black!255, scale=.8]
\newcommand{\roundlabel}[1]{\begin{tikzpicture}[baseline={([yshift=-0.8ex]current bounding box.center)}]\node[roundnode]{\color{white}\textbf{#1}};\end{tikzpicture}}
\tikzstyle{squarenode} = [square, fill=black!255, scale=1.0]

\begin{document}
\fancyhead{}

%
\title{IMAC-Sim: A Circuit-level Simulator For In-Memory Analog Computing Architectures}

\author{Md Hasibul Amin}
\email{ma77@email.sc.edu}
\orcid{0000-0002-9919-6626}
\affiliation{%
  \institution{University of South Carolina}
  \city{Columbia}
  \state{SC}
  \country{US}
  \postcode{29205}
}

\author{Mohammed E. Elbtity}
\email{elbtity@ieee.org}
\orcid{0000-0002-3282-0076}
\affiliation{%
  \institution{University of South Carolina}
  \city{Columbia}
  \state{SC}
  \country{US}
  \postcode{29205}
}

\author{Ramtin Zand}
\email{ramtin@cse.sc.edu}
\orcid{0000-0002-1786-1152}
\affiliation{%
  \institution{University of South Carolina}
  \city{Columbia}
  \state{SC}
  \country{US}
  \postcode{29205}
}

\renewcommand{\shortauthors}{Md Hasibul Amin, Mohammed E. Elbtity, \& Ramtin Zand}

\begin{abstract}
With the increased attention to memristive-based in-memory analog computing (IMAC) architectures as an alternative for energy-hungry computer systems for machine learning applications, a tool that enables exploring their device- and circuit-level design space can significantly boost the research and development in this area. Thus, in this paper, we develop IMAC-Sim, a circuit-level simulator for the design space exploration of IMAC architectures. IMAC-Sim is a Python-based simulation framework, which creates the SPICE netlist of the IMAC circuit based on various device- and circuit-level hyperparameters selected by the user, and automatically evaluates the accuracy, power consumption, and latency of the developed circuit using a user-specified dataset. Moreover, IMAC-Sim simulates the interconnect parasitic resistance and capacitance in the IMAC architectures and is also equipped with horizontal and vertical partitioning techniques to surmount these reliability challenges. IMAC-Sim is a flexible tool that supports a broad range of device- and circuit-level hyperparameters. In this paper, we perform controlled experiments to exhibit some of the important capabilities of the IMAC-Sim, while the entirety of its features is available for researchers via an open-source tool\footnote{\url{https://github.com/iCAS-Lab/IMAC-Sim}}.

\end{abstract}






%

\maketitle

\section{\textbf{Introduction}}

The computational demands of machine learning (ML) workloads are growing rapidly with which the conventional von Neumann architectures cannot keep up  \cite{Wang2019BenchmarkingTG} due to the expensive cost of data transfer between the processor and memory. In recent years, there has been a great interest to use in-memory computing (IMC) as an alternative approach that enables performing the computations directly where the data exist \cite{Sebastian2020MemoryDA}. Emerging memristive devices have been widely used within IMC architectures to realize matrix-vector multiplication (MVM) operations in ML models. While memristive technologies are leveraged in both digital \cite{IMC-Anand-roy} and analog \cite{IMAC,PUMA} IMC circuits, in this work we are focusing on in-memory analog computing (IMAC) architectures due to their great potential for achieving outstanding energy efficiency. For instance, Imec has recently provided a blueprint towards 10,000 tera operations per second per watt (TOPS/W) ML inference in \cite{imec2}, which is based on the memristive-based IMAC architectures.  

\begin{table*}[]
\caption{Comparison of IMAC-Sim with other Circuit-Level Simulation Frameworks}
\vspace{-2mm}
\label{tab:comparison_table}
\centering
\begin{tabular}{lccccccccccc}
\hline
& \multicolumn{4}{c}{Reliability} && \multicolumn{1}{c}{Scalability} && \multicolumn{3}{c}{Output} \\\cline{2-5} \cline{7-7} \cline{9-11}
& Parasitics & Noise & PV & Temperature && Partitioning && Accuracy & Power & Latency \\\hline
DPE \cite{dpengine} & \ding{51} & \ding{51} & \ding{51} & \ding{51} && \ding{53} && Circuit Sim. & \multicolumn{2}{c}{Analytical Model} \\
CCCS \cite{cccs} & \ding{51} &\ding{53} &\ding{53} & \ding{53} && \ding{53} && Circuit Sim. & \ding{53} & \ding{53} \\\hline
\rowcolor[HTML]{EFEFEF} \textbf{IMAC-Sim} & \ding{51} & \ding{51} & \ding{51} & \ding{53} && \ding{51} && \multicolumn{3}{c}{\textbf{Full Circuit Simulation}} \\
\hline
\end{tabular}
\end{table*}

While the interest in IMAC architectures rises, simulation frameworks are required for exploring the design options and tuning the hyperparameters to achieve specific design objectives. The existing IMC simulation frameworks can be classified into two categories: architecture-level \cite{mnsim,neurosim} and circuit-level \cite{dpengine,cccs} frameworks. Architecture-level simulation frameworks use behavioral models and analytical calculations to obtain the characteristics of various IMC architectures, while circuit-level simulators leverage full circuit analysis techniques to analyze the functionality and performance of the IMC circuits. Though architecture-level tools are faster than circuit-level frameworks, they typically exclude some important design details which can lead to important discrepancies between simulation and real implementation. On the other hand, full circuit  simulations can provide accurate results at the cost of increased simulation time. 

Two of the well-known architecture-level simulation frameworks are MNSim \cite{mnsim} and NeuroSim \cite{neurosim}, which use analytical models to estimate the area, power, and latency of IMC architectures. The MNSim validation results show roughly 5\% error in power, energy, and latency calculation compared to a SPICE circuit simulation for a two-layer fully connected neural network. While, MNSim does not report actual accuracy results, NeuroSim \cite{neurosim} is integrated with ML simulators to obtain the learning and classification accuracy values but it is still missing the accurate circuit-level prediction model for the analog behavior of the IMC arrays.

DPE \cite{dpengine} and CCCS \cite{cccs} are two of the main circuit-level simulation platforms for IMC circuits. Both frameworks use circuit-level solvers developed in MATLAB. The main focus of DPE \cite{dpengine} is on creating an optimized scheme for mapping the pre-trained weights to memristive crossbars considering the non-ideal effects. Instead of calculating the interconnect parasitics based on parameters such as size of the array, resistivity, interconnect thickness, etc., DPE takes a fixed predefined value as input for parasitic simulation. Also, DPE gets the power and latency of the building blocks and then estimates the power and latency of the entire architecture using analytical calculations, hence similar simulation accuracy losses as the architecture-level simulators are expected to be observed. On the other hand, CCCS \cite{cccs} framework uses model simplification techniques for evaluating the parasitic effect on the crossbar outputs by dividing the crossbar into small sub-blocks in which the parasitic effects are ignored. While creating sub-blocks reduces the number of nodes that are required to be processed simultaneously and reduces simulation time, it leads to increased calculation error due to inaccurate estimation of interconnect voltage drop. CCCS validation results \cite{cccs} show that the simplification methods can achieve almost 30$\times$ simulation speedup at the cost of up to 27\% simulation error which can have severe impacts on the design and optimization of IMAC circuits. In addition, CCCS \cite{cccs} framework only provides ML model accuracy results and does not support power and latency calculations.   

In this paper, we propose the IMAC-Sim framework for accurate SPICE-level modeling of in-memory analog computing architectures. IMAC-Sim is a highly-modular and parameterized tool that is capable of reporting the accuracy, power consumption and latency by means of a full circuit SPICE simulation.  Table \ref{tab:comparison_table} provides a comparison between IMAC-Sim and previous circuit-level IMC simulators. As listed, besides parasitics, DPE simulates noise, process variation (PV), and temperature effects for memristor devices while CCCS considers none of them. So far, IMAC-Sim has the capability to investigate noise and PV effects. The unique features of IMAC-Sim include, (1) automatic calculation of parasitic resistances and capacitances based on layout design information and interconnect parameters such as resistivity, interconnect width, thickness, etc, (2) implementation of partitioning techniques to distribute ML workloads on partitioned subarrays and thus decreasing the effect of interconnect parasitic on large-scale networks, and (3) accurate power and latency report through full circuit SPICE simulation.

\section{\textbf{Proposed IMAC-Sim Framework}}

\begin{figure}[t]
\centering
\includegraphics[width=3.4in]{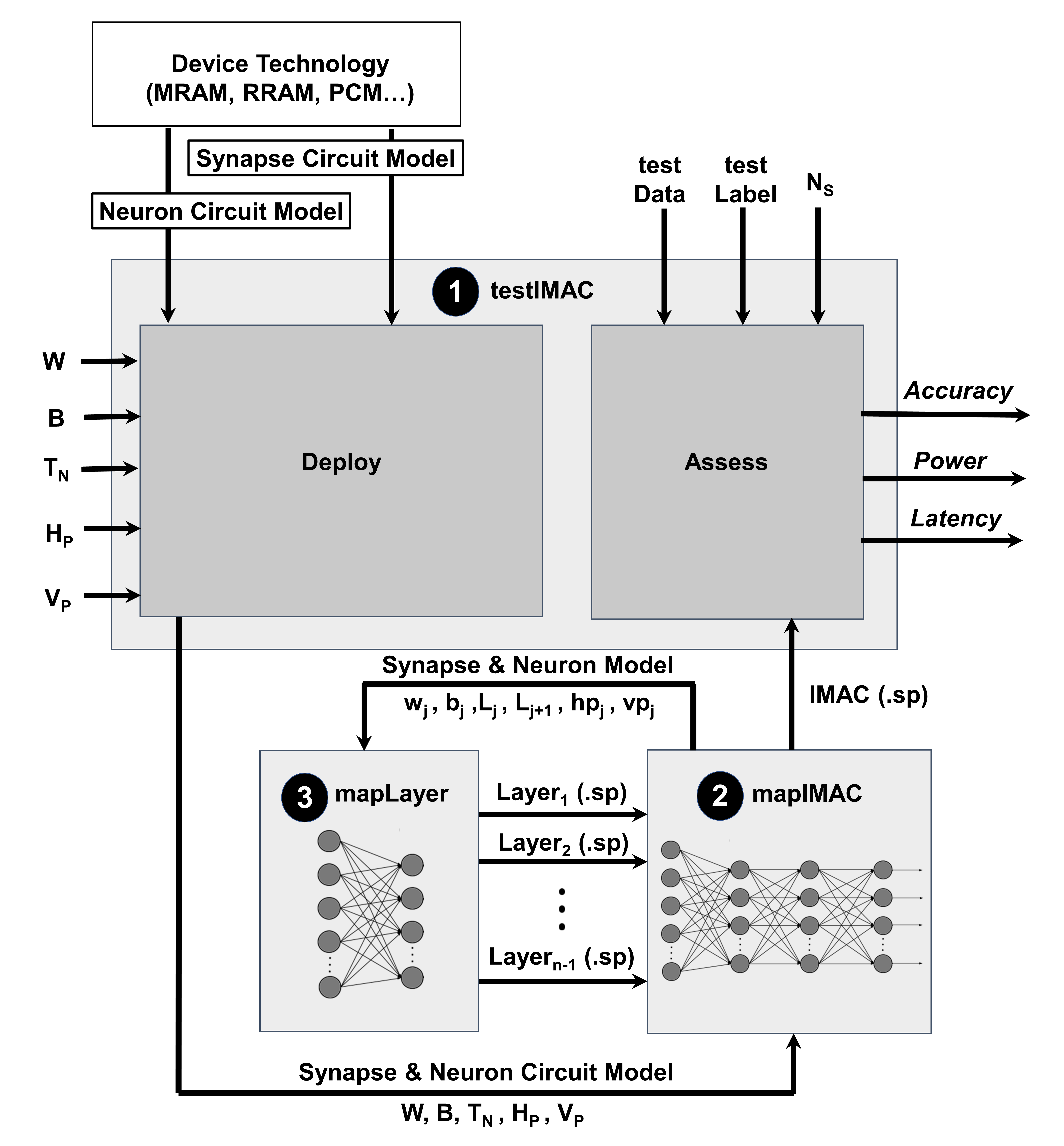}
\vspace{-6mm}
\caption{Block diagram of the IMAC-Sim framework.}
\label{fig:imacflow}
\vspace{-5mm}
\end{figure}

Figure \ref{fig:imacflow} shows the block diagram of the IMAC-Sim framework which generates the SPICE netlist of IMAC circuits based on the network topology and specific user-defined device- and circuit-level hyperparameters listed in Table \ref{tab:hyperparameters}. After generating the IMAC circuits, IMAC-Sim automatically runs various ML tasks, the corresponding datasets of which are provided by the user, on the developed IMAC circuit by running the SPICE simulation. It extracts the outputs of the circuit and compares them with the target labels to evaluate the IMAC accuracy. Moreover, it measures the power consumption and latency and provides them as an output to the user along with the calculated accuracy. The users can leverage this information to optimize their developed circuits. IMAC-Sim framework includes three main Python modules: \roundlabel{1} \textit{testIMAC}, 
\roundlabel{2} \textit{mapIMAC}, and \roundlabel{3} \textit{mapLayer}, which are shown in Fig. \ref{fig:imacflow} and described in the following.

\begin{table}[]
\caption{IMAC-Sim hyperparameters.}
\vspace{-2mm}
\label{tab:hyperparameters}
\centering
\begin{tabular}{lc}
\hline
Parameter               & Value    \\ \hline

Network Topology & $T_N=[L_1, L_2, ...,L_n]$ \\
Transistor Technology Node             & FinFET, CMOS     \\
Nominal Voltages         & [\textit{VDD}, \textit{VSS}]     \\
Neuron Circuit Model            & \textit{sigmoid}, \textit{tanh}, ReLU, \textit{etc.}          \\
Synaptic Technology        & [$R_{low}$ , $R_{high}$]       \\
Vertical Partitioning   & $V_P$ = [$vp_1$, $vp_2$,..., $vp_{n-1}$]  \\
Horizontal Partitioning & $H_P$ = [$hp_1$, $hp_2$,..., $hp_{n-1}$]   \\
Synapse Bitcell Size                     &     [\textit{Width}, \textit{Height}]             \\
Interconnect           & [\textit{resistivity}, \textit{thickness}, \textit{width}, ...]        \\
Sampling Time & $t_{sampling}$
\\

\hline
\end{tabular}
\vspace{-5mm}
\end{table}

The \textit{testIMAC} module runs as a parent file, which controls the deployment of deep neural network (DNN) workloads on IMAC architectures, as well as assessing the accuracy, power, and latency. The \textit{testIMAC} module receives two sets of inputs from the user; one for deployment and another set for assessment. For deployment, it receives pre-trained $weight$ and $bias$ matrices, network topology ($T_N$), number of horizontal partitions per layer ($H_P$) and number of vertical partitions per layer ($V_P$), and the device- and circuit-level hyperparameters to deploy the DNN on the IMAC architecture. For assessment, it receives the test dataset ($TestData$), test label ($TestLabel$), and the number of test samples ($N_S$) to assess the developed IMAC circuit. $T_N$ is an array of data containing the number of nodes from input to output layers in DNN. $H_P$ and $V_P$ are arrays, in which the number of horizontal and vertical partitions for each layer is stored. The functionality of \textit{testIMAC} module is demonstrated in Algorithm \ref{algo:testIMAC}.

\begin{algorithm}
\DontPrintSemicolon
\SetAlgoLined
\nonl $/*\ Module\ \roundlabel{1}$\;
\KwIn{test dataset ($TestData$), test label ($TestLabel$), weights ($W$), biases ($B$), Network Topology ($T_N$), No. of test samples ($N_S$), Horizontal Partitioning ($H_P$), Vertical Partitioning ($V_P$), $R_{low}$, $R_{high}$}
\BlankLine

 \textbf{Initialize:} $Error=0$, $PWR=0$, $W=[w_1, w_2, ...,w_{n-1}]$, $B=[b_1, b_2, ...,b_{n-1}]$, $T_N=[L_1, L_2, ...,L_n]$, $H_P=[hp_1, hp_2, ...,hp_{n-1}]$, $V_P=[vp_1, vp_2, ...,vp_{n-1}]$\;
 
 \For{$i=1$ \KwTo $N_S$}{
  \textit{label} $\Leftarrow$ target labels for input $i$\;
  \nonl $/*\ Module\ \roundlabel{2}$\;
  \SetKwProg{Fn}{}{:}{\KwRet {\textit{IMAC} SPICE circuit}}
  \Fn{\textbf{mapIMAC}{ ($W$, $B$, $T_N$, $H_P$, $V_P$)}}{  
    
     \For{$j=1$ \KwTo $j=len(T_N)-1$}{
     \nonl $/*\ Module\ \roundlabel{3}$\;
     \SetKwProg{Fn}{}{:}{\KwRet {$Layer_j$ SPICE subcircuit}}
  \Fn{\textbf{mapLayer}{ ($w_j$, $b_j$, $L_j$, $L_{j+1}$ $hp_j$, $vp_j$)}}{
  Implement the partitioning ($hp_j$, $vp_j$)\;
  Insert interconnect parasitics\;
  }}
  Concatenate the $layer_j$ subcircuits\;
  }
  Run the SPICE simulation for \textit{IMAC} circuit\;
  \textit{out} $\Leftarrow$ Output of the IMAC circuit \;
  $PWR \mathrel{{+}{=}}$ power consumption of IMAC circuit\;
  
  \If{(out$\neq$label)}{
  $Error  \mathrel{{+}{=}} 1$;
  }
}
print $ErrorRate=Error/N_S$\;
print $P_{average}=PWR/N_S$\;
 \caption{Implementation of IMAC using \textit{IMAC-Sim} Framework}
 \label{algo:testIMAC}
\end{algorithm}

As shown in Algorithm \ref{algo:testIMAC}, for each input sample, \textit{testIMAC} first stores the target labels in an array called $label$. It then calls another python module, \textit{mapIMAC}, which is responsible for creating the SPICE netlist of the IMAC circuit. For this purpose, \textit{mapIMAC} calls another python module, \textit{mapLayer}, which builds separate subcircuits for each of the layers in DNN including their interconnect parasitics and required partitioning as requested by the user through $H_P$ and $V_P$ arrays. We discuss parasitics and partitioning in detail in the following two subsections. The \textit{mapLayer} modules return the SPICE files for all of the layer subcircuits to \textit{mapIMAC}, which concatenates them to form the main \textit{IMAC} SPICE file. Finally, \textit{testIMAC} runs the SPICE simulation for the developed \textit{IMAC} SPICE file using the input voltages generated from the test dataset, and extracts the outputs of the last layer in IMAC circuit ($out$) and compares them with the $label$ to obtain the accuracy. Moreover, \textit{testIMAC} measures the average power consumption and latency of the circuit across various inputs and reports them to the user.

\subsection{Interconnect Parasitics}
\label{sec:parasitic}
Parasitic interconnect resistance ($R_W$) and capacitance ($C_W$) are a function of wire geometry and material properties of interconnections in IMAC subarrays. Scaling up the size of IMAC arrays increases $R_W$ and $C_W$, which can consequently increase the latency of the IMAC circuits and limit their operating clock frequency. Moreover, increased $R_W$ reduces the IMAC read margin that can impact its accuracy \cite{Aguirre2020,iscas-imac}. The interconnect parasitic resistances for IMAC can be found using the below equation:

\begin{equation}
R_W = \rho\frac{L}{W.T},
\end{equation}

\noindent where $\rho$, $L$, $W$ and $T$ are the resistivity, length, width, and thickness of the metal wire, respectively. While resistivity is commonly considered a fixed parameter  for a specific metal, for the below 10nm technology nodes where the metal width is near the mean free path of electrons, the resistivity increases due to the surface and grain boundary scattering \cite{doi:10.1146/annurev-matsci-082908-145415}. These two well-known scattering effects can be quantified using Fuchs-Sondheimer (FS) \cite{fuchs_1938} and Mayadas-Shatzkes (MS) \cite{PhysRevB.1.1382} models, as described in the following,

\begin{equation}
    \frac{\rho}{\rho_{Cu}}=\frac{3}{4}(1-p)\frac{l_0}{W}+\left[1-\frac{3\alpha}{2}+3\alpha^2-3\alpha^3\ln\left(1+\frac{1}{\alpha}\right)\right]^{-1}\\
\end{equation}

\noindent where $\alpha=\frac{l_0}{d}\frac{R}{1-R}$, $\rho_{Cu}$ is the resistivity of bulk Cu ($1.9\times10^{-9}$ $\Omega m$), $l_0$ is the mean free path of electrons in Cu (39 nm), $W$ is the width of the metal interconnect, $p$ is the specular scattering fraction, $d$ is the average grain size and $R$ is the probability for electrons to reflect at the grain boundary. $R$ and $p$ are assumed to be 0.3 and 0.25, respectively, and $d$ is assumed to be equal to the interconnect width, based on the average values mentioned in the literature \cite{rossnagel},\cite{steinhogl}.

The parasitic capacitances also play a major role in determining the latency of IMAC. 
Here, we use the Sakurai-Tamaru model \cite{1482994} for calculating the parasitic capacitance per length, as described in the below equation,

\begin{equation}
\begin{aligned}
C_W =&\frac{\epsilon}{2}\left[1.15\left(\frac{W}{H}\right)+2.8\left(\frac{W}{H}\right)^{0.222}\right]+2\epsilon\left(\frac{S}{H}\right)^{-1.34} \\
& \times \left[0.03\left(\frac{W}{H}\right)+0.83\left(\frac{T}{H}\right)-0.07\left(\frac{T}{H}\right)^{0.222}\right]
\end{aligned}
\end{equation}

\noindent where $\epsilon=20\epsilon_0$ is the dielectric permittivity of the inter-metal space, $W$ and $T$ are the width and thickness of the metal line, $H=20nm$ is the inter-metal layer spacing and $S$ is the inter-wire spacing \cite{Aguirre2020}. IMAC-Sim leverages equations (2) and (3) to calculate and inject the interconnect parasitics into the IMAC circuits.

\subsection{Horizontal and Vertical Partitioning}

\begin{algorithm}[t]
\DontPrintSemicolon
\SetAlgoLined
\KwIn{weights ($w$), biases ($b$), Layer Dimension ($L_1$, $L_2$), No. of Horizontal Partitions ($h_p$), No. of Vertical Partitions ($v_p$)}
\BlankLine

 
 \For{$i=1$ \KwTo $h_p$}{
    \textit{$NH_{i}$} $\Leftarrow$ $L_1+1$ \;
        \textit{$NHP_{i}$} $\Leftarrow$ $NH_{i}\times i/h_p+min((NH_{i} \mod h_p)/i,1)$ \;
     \For{$j=1$ \KwTo $v_p$}{
        \textit{$NVP_{j}$} $\Leftarrow$ $L_2\times j/v_p+min((L_2\mod v_p)/j,1)$ \;
        \nonl $/*\ Partition$ $P_{ij}$\;
        Place a 2D matrix of bitcells with dimension $NHP_{i} \times NVP_{j}$ \;
  }
  }
  
 \caption{Partitioning ($h_p$,$v_p$)}
 \label{algo:part}
\end{algorithm}

As the interconnect parasitics can severely degrade the accuracy of the IMAC circuits, horizontal and vertical partitioning techniques have been proposed in the literature \cite{xbar,Aguirre2020} to decrease $R_W$ and $C_W$ in the IMAC circuits. We have equipped the IMAC-Sim with both horizontal and vertical partitioning, which we describe in algorithm \ref{algo:part}. The number of horizontal and vertical partitions for each layer in DNN is a hyper-parameter that can be tuned by the user through $H_P$ and $V_P$ arrays in the \textit{testIMAC} module, respectively. The \textit{testIMAC} module calls the $partitioning (h_p,v_p)$ for each of the layers separately, which automatically partitions that layer and connects the peripheral circuits. As shown in algorithm \ref{algo:part}, the partitioning function iterates over $h_p$ and $v_p$, and connect all the partitions one by one. $NH_{i}$ represents number of bitcells in the horizontal direction. $NHP_{i}$ and $NVP_{j}$ represents the number of bitcells for each of the partitions in horizontal and vertical directions, respectively. Thus, a 2D matrix of bitcells with dimension $NHP_{i} \times NVP_{j}$ is connected for the partition $P_{ij}$.

\section{\textbf{Simulation Results and Discussion}}

In this section, we utilize our developed IMAC-Sim framework to implement a $400\times120\times84\times10$ DNN for classification application using an MNIST dataset with 20$\times$20 pixel handwritten digit images. It shall be emphasized that our focus here is not on designing an optimized IMAC architecture, and instead we are exhibiting some of the important features and capabilities of IMAC-Sim tool. Thus, developing various DNN models on IMAC architectures for different types of datasets is out of the scope of this paper. That said, we invite researchers to leverage our open-source tool that is available at the project's GitHub repository 
for benchmarking their IMAC designs and guiding early-stage design decisions. 

\subsection{Constraining the Design Space of Experiments}
Herein, to limit the wide IMAC design space, we have fixed some of the hyperparameters as listed in Table \ref{tab:hyperpardefaults}, while the impact of the other parameters on accuracy and power consumption of IMAC is investigated in the following subsection. We use the 14nm High-Performance PTM-MG FinFET model along with the $VDD$ and $VSS$ voltages of 0.8V and -0.8V, respectively. Based on the 18nm gate length and the 22nm Fin height of the PTM 14nm FinFET model \cite{ptmpaper}, the layout design parameter $\lambda$ and the metal thickness are fixed to 9nm and 22nm, respectively.

\begin{table}[]
\caption{Hyper-parameters fixed herein to limit the IMAC design space.}
\vspace{-2mm}
\label{tab:hyperpardefaults}
\centering
\begin{tabular}{lc}
\hline
Parameter               & Value    \\ \hline

Transistor Technology Node             & 14 nm FinFET    \\ \hline
Nominal Voltages         & [\textit{VDD} $=0.8$V, \textit{VSS}$=-0.8$V]     \\ \hline
Neuron Circuit Model            & Sigmoid         \\ \hline


\multirow{3}{*}{Interconnect} &  \textit{Resistivity $\rho=1.9\times 10^9 \Omega.m$}  \\
                   & \textit{Thickness$=22  nm$} \\
                   & \textit{Width$=4\lambda= 36 nm$} \\ \hline

Inter-metal layer spacing & $20 nm$
\\

\hline
\end{tabular}
\vspace{-5mm}
\end{table}


 We use an analog sigmoidal neuron that includes two resistive devices and a CMOS-based inverter. The resistive devices in the neuron's circuit create a voltage divider that reduces the slope of the linear operating region in the inverter leading to a smooth high-to-low output voltage transition, which enables the realization of a $sigmoid$ activation function \cite{glsvlsi_neuron}. Users can utilize any other neuron circuit through creating a SPICE subcircuit file called ``neuron.sp'' and placing it within the same directory where IMAC-Sim tool is stored. By fixing the interconnect parameters, as well as synapse and neuron structures, we constrain the wide design space of the IMAC to focus on some of the interesting features of IMAC-Sim as described in the following.

\subsection{Design Space Exploration}

\begin{figure}[t]
    \begin{subfigure}{.5\textwidth}
        \centering
        \includegraphics[width=3.4in]{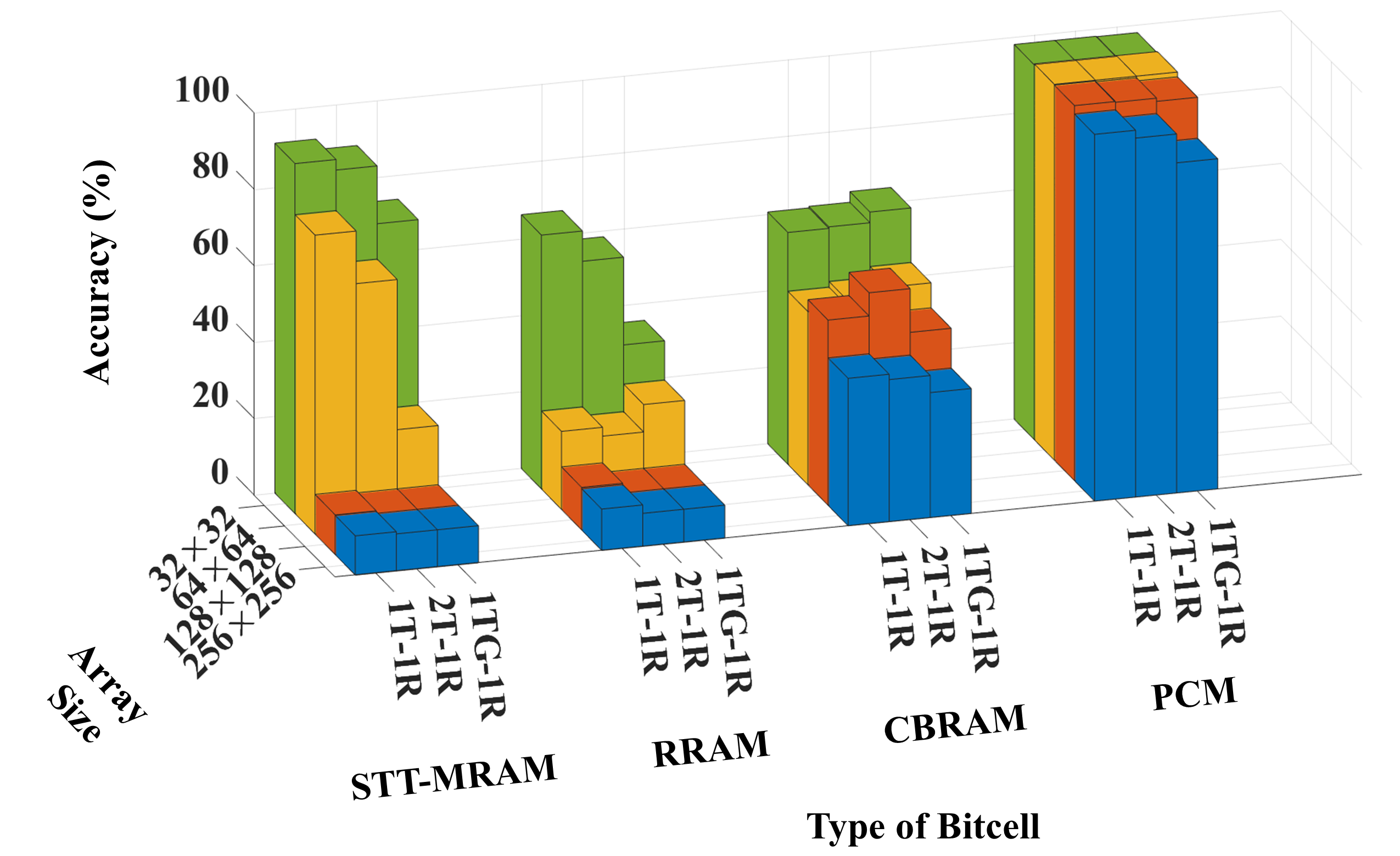}
        \caption{}
        \label{fig:accuracy_bar}
    \end{subfigure}
    \begin{subfigure}{.5\textwidth}
        \centering
        \includegraphics[width=3.4in]{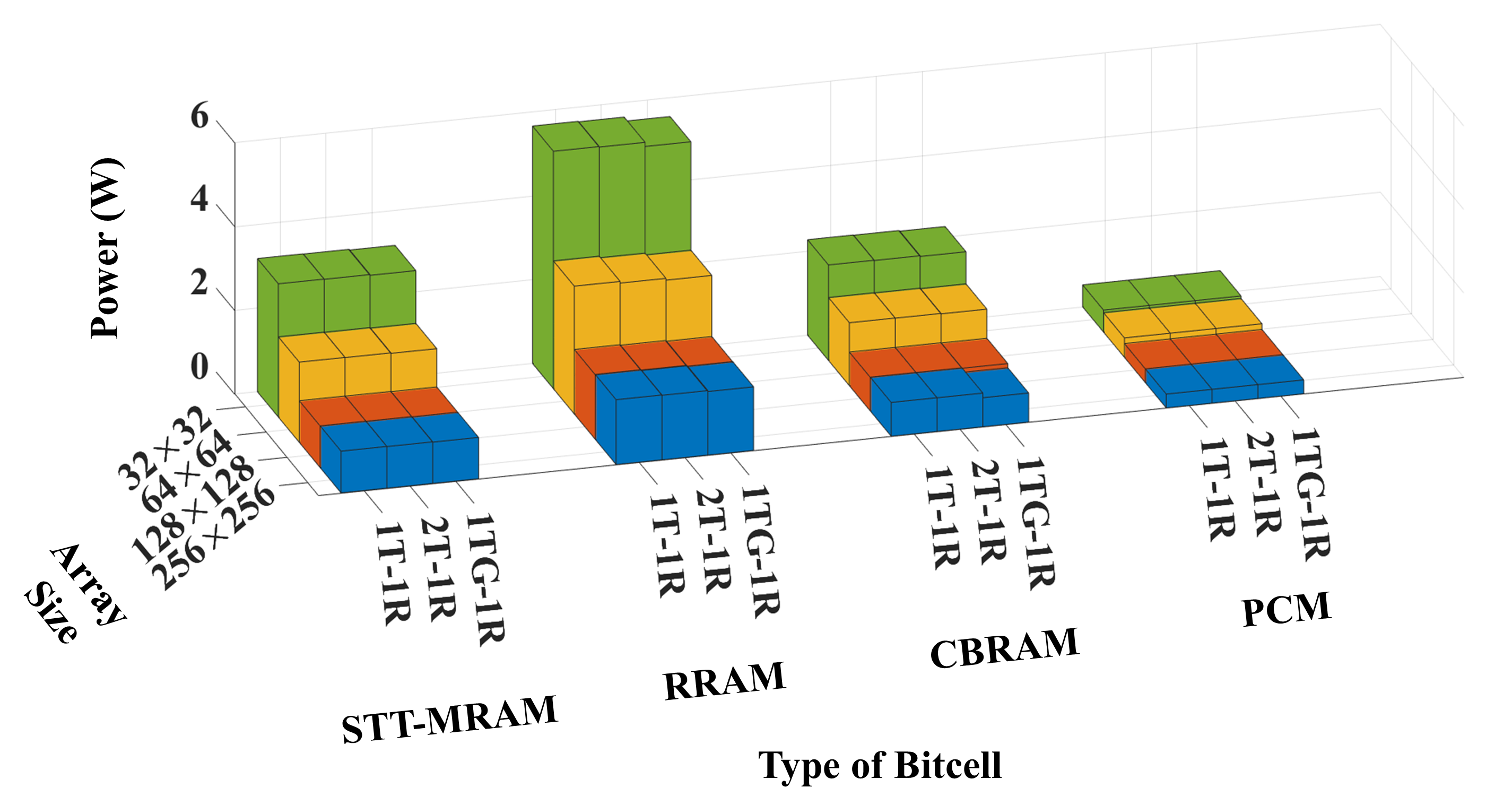}
        \caption{}
        \label{fig:power_bar}
    \end{subfigure}
    \vspace{-5mm}
    \caption{ The results obtained for the deployment of a $400\times120\times84\times10$ DNN model on IMAC with various subarray sizes, memristive technologies, and bitcell types. (a) Accuracy and (b) Power consumption.}
    \vspace{-5mm}
    \label{fig:bar}
\end{figure}

In this section, we use IMAC-Sim to explore the design space of IMAC with the $400\times120\times84\times10$ DNN model for various device technologies, synapse bitcell size and no. of partitions. The synapse bitcell size is an important hyperparameter to estimate the distance between the interconnects and find the length of the metal lines, which impacts the interconnect parasitic resistance and capacitance as described in Section \ref{sec:parasitic}. We explore the results with three different types of bitcells 1) 1T-1R (Length: $15\lambda$, Width: $12\lambda$), 2) 2T-1R (Length: $15\lambda$, Width: $12\lambda$) \cite{layout-aware} , and 3) 1TG-1R (Length: $15\lambda$, Width: $12\lambda$) \cite{zandTVLSI}. Also, we run experiments with four different device technologies 1) magnetoresistive random-access memory (MRAM) ($R_{low}: 8.5k\Omega$, $R_{high}: 25.5k\Omega$) \cite{zand2018fundamentals}, 2) Resistive random access memory (RRAM) ($R_{low}: 2.5k\Omega$, $R_{high}: 100k\Omega$) \cite{li2018analogue}, 3) conductive bridging random access memory (CBRAM) ($R_{low}: 10k\Omega$, $R_{high}: 1M\Omega$) \cite{shi2018neuroinspired}, and 4) phase-change memory (PCM) ($R_{low}: 78k\Omega$, $R_{high}: 202k\Omega$) \cite{pcm_new}. 

Also, we study the impact of partitioning by varying the number of horizontal and vertical partitions. While the IMAC-Sim supports any arbitrary value for horizontal and vertical partitioning, here we have selected the number of partitions for each layer based on the maximum utilization of IMAC subarrays with various dimensions. For instance, if we use $256 \times 256$ subarrays, the first layer that includes 400 inputs must be divided into two horizontal partitions to fit into two $256 \times 256$ IMAC subarrays, while the other layers can fit into the arrays without any horizontal and vertical partitioning. We run IMAC-Sim with each partitioning scenario for various device types and bitcell sizes to measure the accuracy and power consumption, which are shown in Fig. \ref{fig:bar}. The results shown in Fig. \ref{fig:bar} demonstrate that as the number of horizontal and vertical partitions increases in the smaller-sized IMAC subarrays, the accuracy and power consumption increase as well. For instance, highly partitioned deployment of the DNN model on $32 \times 32$ IMAC subarrays provides 91.6\% accuracy on the MRAM-based 1T-1R bitcell, while the accuracy is just 11\% with lower partitions on $256 \times 256$ similar type of subarrays. However, power consumption with lower partitions is 0.996W while it is as high as 3.18W with higher partitions in $32 \times 32$ IMAC subarrays. Also, if we consider 1TG-1R bitcells, the accuracy drops to 73.72\% from 91.6\% as mentioned above, which shows that IMAC is more affected by the parasitics if the bitcell is larger. Also, for PCM-based bitcell with a similar arrangement, the accuracy improves to 98.56\% with a power consumption of 0.52W.

\begin{figure}[t]
\centering
\includegraphics[width=3.4in]{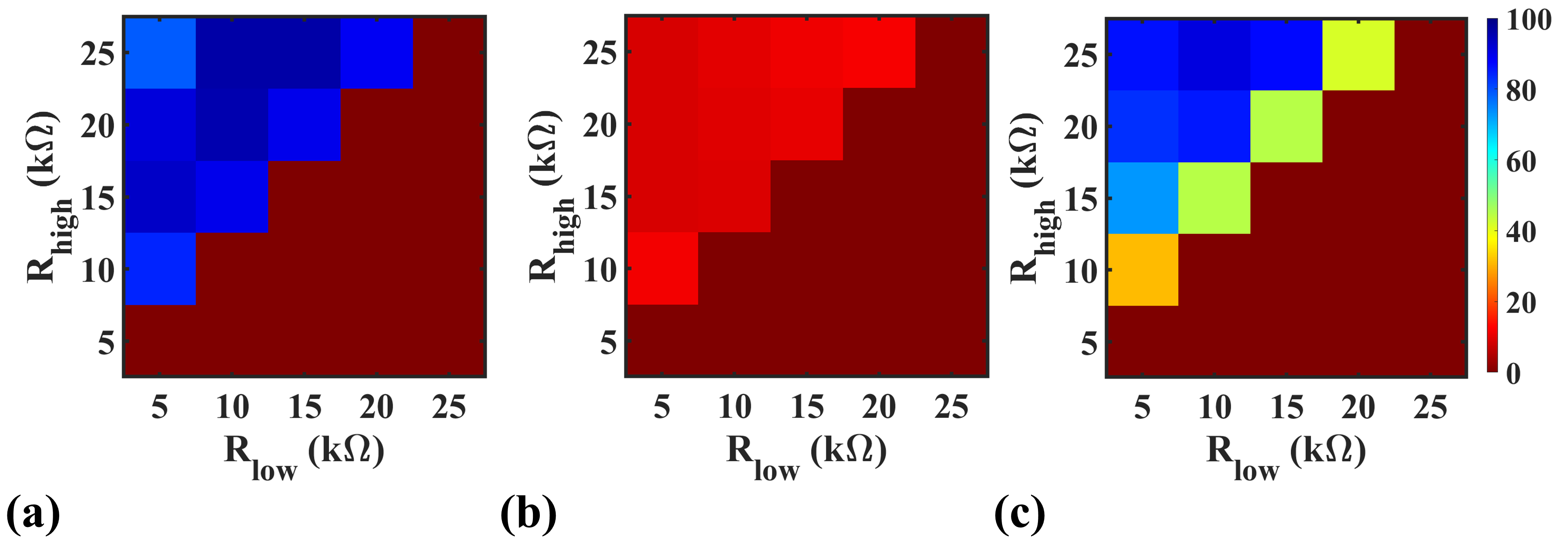}
\vspace{-7mm}
\caption{Accuracy results obtained from IMAC-Sim (a) without parasitics, (b) with parasitics, and (c) with parasitics and partitioning.}
\label{fig:accuracy}
\vspace{-5mm}
\end{figure}

To investigate the impact of device technologies on the performance of the IMAC architecture, we sweep the $R_{low}$ and $R_{high}$ values and obtain the DNN accuracies from IMAC-Sim considering three different scenarios, which are shown in Fig. \ref{fig:accuracy}. The accuracy values on the bottom right side of the figures are always equal to zero because the circuit fails when $R_{high}$ is less than or equal to $R_{low}$. The model shows high accuracy values in Fig \ref{fig:accuracy}a where we ignore the effect of parasitic resistances. Then we add parasitics to the circuit, which significantly affects the DNN accuracy, as shown in Fig. \ref{fig:accuracy}b. Fig. \ref{fig:accuracy}c shows the results for the partitioned deployment of the DNN model on $32 \times 32$ IMAC subarrays. It is clear that partitioning improves the accuracy, which was severely degraded by the parasitic effects. As an example, with $R_{low}=5k \Omega$ and $R_{high}=15k \Omega$, the accuracy without parasitics is 93.3\%, while it drops to 8.9\% with parasitics, then partitioning improves it to 72.7\%. Fig. \ref{fig:accuracy} shows that the accuracy improvement by partitioning highly depends on the $R_{low}$ and $R_{high}$ values. Thus, IMAC-Sim will enable the designers to take early stage decisions on their choice of device technology.

\begin{figure}[t]
\centering
\includegraphics[width=3.4in]{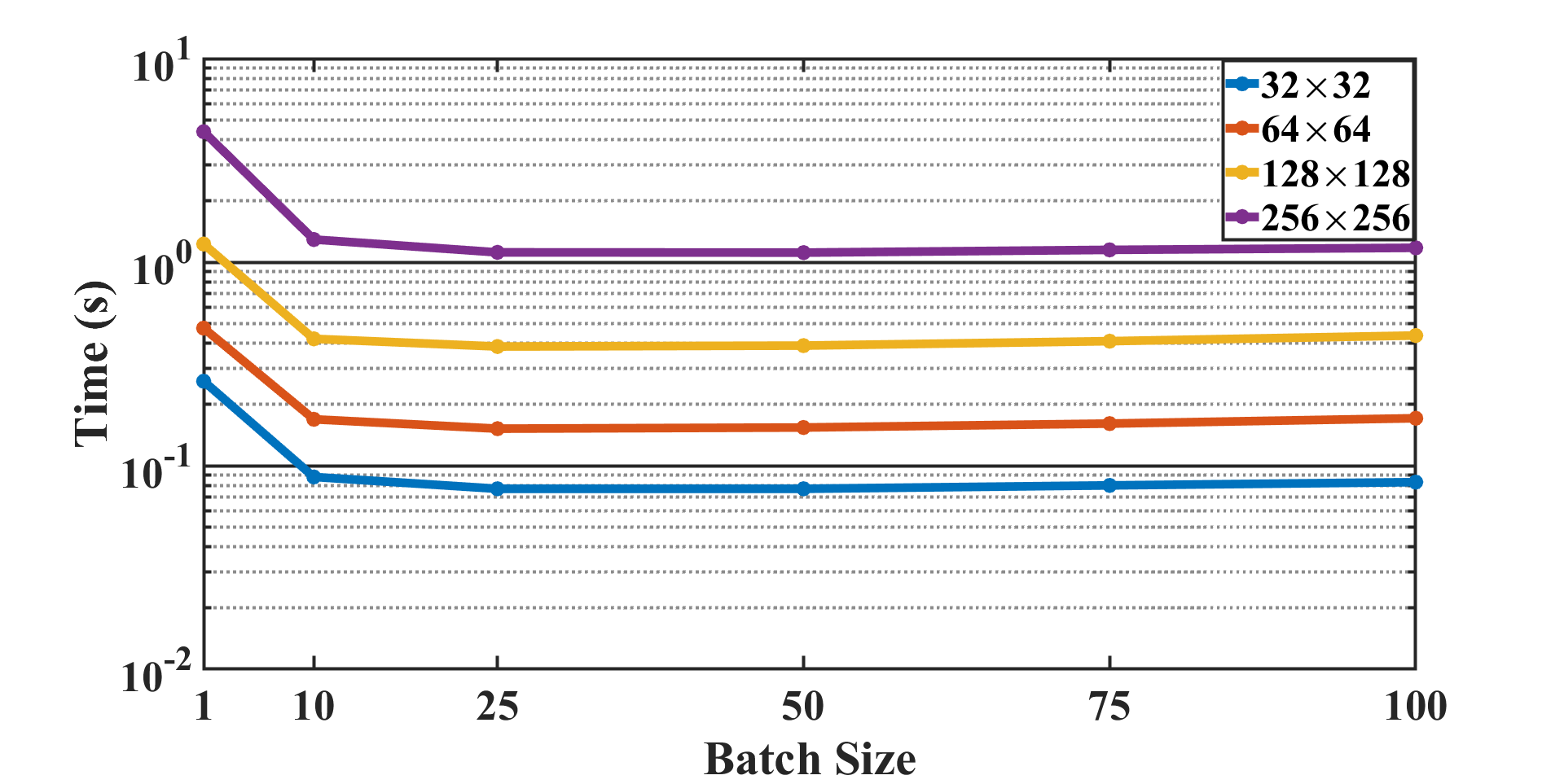}
\vspace{-6mm}
\caption{Batch size vs average simulation time for various array sizes.}
\label{fig:time}
\vspace{-5mm}
\end{figure}

\subsection{Simulation Time Analysis}
In this section, we run the IMAC-Sim framework for various IMAC subarrays on an Intel Xeon Silver 4210R CPU and collect the simulation time information. To speed up the simulation time, we enable IMAC-Sim to run multiple tests in each SPICE simulation by concatenating the input voltages from multiple tests into one piecewise linear (PWL) function. Here, we define a ``batch size'' parameter which represents the number of input tests, e.g., images for MNIST, that are concatenated for each simulation. To implement batches in SPICE, we use the PWL signal in which the voltage level is changed every 1ns based on the test case. For example, a batch size of 50 leads to an input waveform with a 50ns length. The 1ns is also the sampling interval for recording the output.  

\begin{figure}[t]
\centering
\includegraphics[width=3.4in]{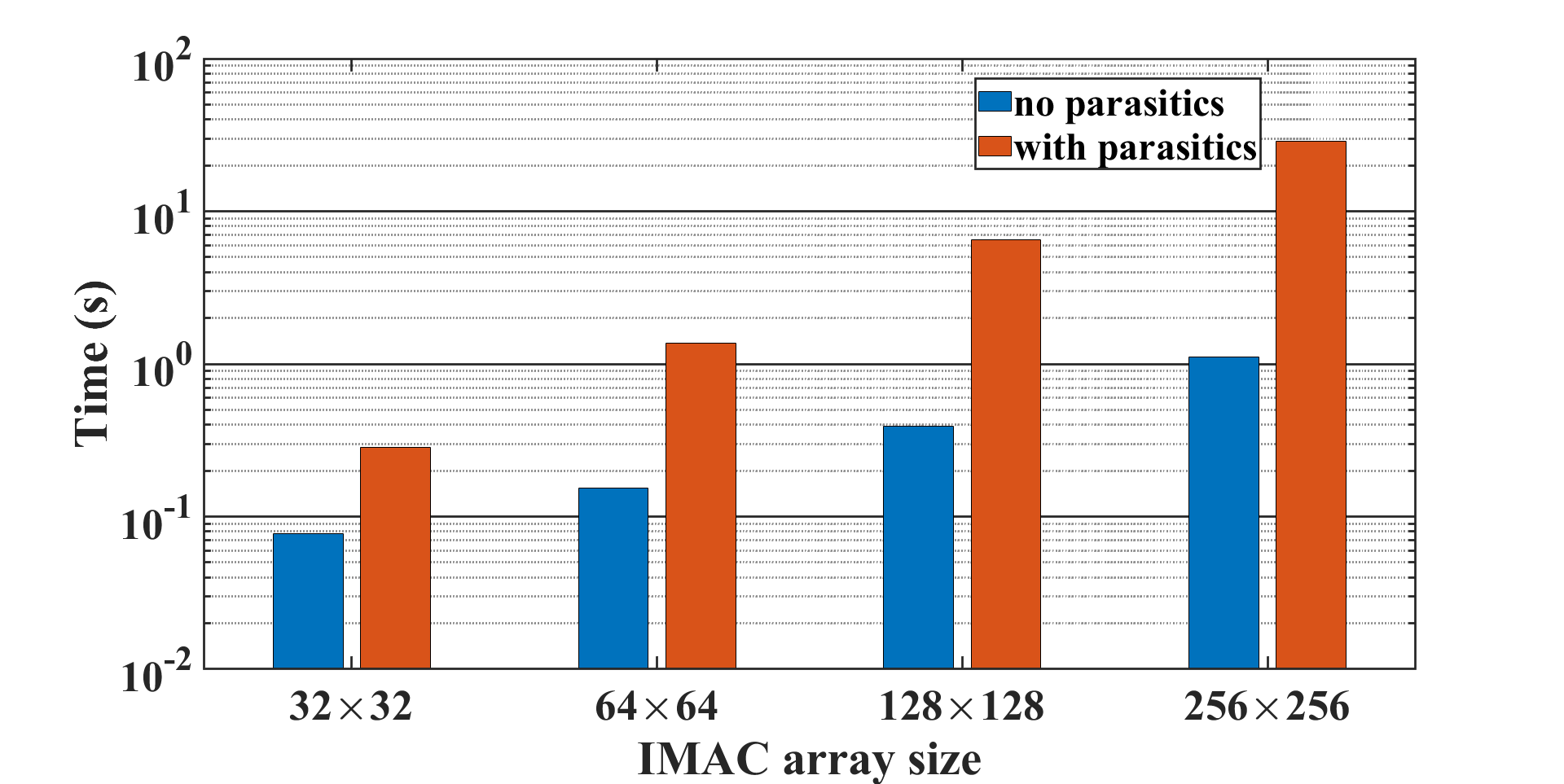}
\vspace{-5mm}
\caption{Average simulation time with and without parasitics for various array sizes for a batch size of 50.}
\label{fig:ptime}
\vspace{-5mm}
\end{figure}

In Fig. \ref{fig:time}, we show the average simulation time for different batch sizes while running various-sized IMAC subarrays without including the interconnect parasitics. The results show that the simulation time is minimized at a specific batch size. For instance, the average time required for the simulation of one $256 \times 256$ IMAC subarray is 4.37 seconds when the batch size is one. The average simulation time drops to 1.11 seconds with a batch size of 50, while it increases to 1.18 seconds for a batch size of 100. Fig. \ref{fig:ptime} shows the simulation time with and without parasitics for various array sizes with batch size 50. The results show that circuit simulation with interconnect parasitics can lead to approximately one order of magnitude increase in the simulation time. Based on the results shown in Fig. \ref{fig:ptime}, the average simulation time for a $256 \times 256$ IMAC subarray with parasitics is 28.76 seconds, while it is just 1.11 seconds without parasitics.

\section{\textbf{Conclusion}}
In this paper, we introduced IMAC-Sim, which is a Python-based simulation framework for SPICE circuit realization of IMAC architectures. IMAC-Sim is a flexible tool that supports a broad range of device- and circuit-level hyperparameters for designing memristive-based IMAC architectures. Some of the distinguishing characteristics of IMAC-Sim compared to previous circuit simulators for in-memory computing architectures include providing a DNN accuracy report to the users, including reliability challenges such as interconnect parasitics, and providing solutions like partitioning to surmount these challenges. Considering the wide IMAC design space supported by IMAC-Sim, we performed experiments to exhibit some of the unique benefits of the IMAC-Sim by constraining the design space via fixing some of the hyperparameters. Our experiments included investigating the impact of partitioning and various memristive device technologies on the accuracy and power consumption of the IMAC architectures. The results obtained by IMAC-Sim included some of the important tradeoffs that can be provided to the developers at early design stages, which showcased some of the potential benefits of IMAC-Sim for circuit designers and computer architects. The development of the IMAC-Sim provides several opportunities for future work including integrating IMAC-Sim with multi-objective optimization algorithms to realize optimized IMAC designs for various ML workloads.

\vspace{-1mm}

\printbibliography


%



\end{document}